\documentclass[nofootinbib,aps,twocolumn]{revtex4}
\usepackage{epsf}

\def\barD{\overline D{}^0}
\def\DDbar{D{}^0-\overline D{}^0}

\def\beq{\begin{equation}}
\def\eeq{\end{equation}}
\def\bea{\begin{eqnarray}}
\def\eea{\end{eqnarray}}

\begin{document}

\preprint{\vbox{\hbox{WSU--HEP--0402}} \hbox{MCTP-04-10}}

\vspace*{2cm}

\title{\boldmath Hunting for CP violation with untagged charm decays} 

\author{Alexey A.\ Petrov\vspace{8pt}}
\affiliation{Department of Physics and Astronomy,
	Wayne State University, Detroit, MI 48201}
\affiliation{Michigan Center for Theoretical Physics, 
University of Michigan, Ann Arbor, MI 48109}

\begin{abstract}
We construct a CP-violating observable which does not require flavor or 
CP tagging of the initial state. The proposed decay asymmetry
could be measured at both threshold and non-threshold charm physics
experiments and provide better sensitivity to small CP-violating
parameters in charm decays.
\end{abstract}
\maketitle

%%%%%%%%%%%%%%%%%%%%%%%%%%%%%%%%%%%%%%%%%%%%%%
\section{Introduction}

One of the most important motivations for studies of
CP violation in charm decays is the possibility of observing 
signals of new physics. It is expected that
CP-violating amplitudes generated by Standard 
Model (SM) interactions are numerically small~\cite{Petrov:2003un},
so observation of CP-violation in the current round of experiments 
will provide a ``smoking gun'' signal of new physics, even if its 
source is not clearly identified. This robust prediction follows 
from the fact that weak decay of the charmed meson or baryon is 
governed by the real $2\times 2$ Cabbibo matrix, as all quarks 
which build up the hadronic states in non-leptonic weak decay 
belong to the first two generations. The only possible SM CP-violating
amplitude comes from the virtual $b$ quarks in penguin or box-type diagrams,
but it is strongly suppressed by the the small combination of 
Cabbibo-Kobayashi-Maskawa (CKM) matrix elements 
$V_{cb}^* V_{ub}\sim {\cal O}(\lambda^5)$, where $\lambda=0.2$ is the 
Wolfenstein parameter. This small ``Standard Model background'' makes charm 
decays a valuable tool in searching for CP-violating effects of new physics, 
especially since the acquired datasets available in charm physics experiments 
are usually quite large.

Let us first review the relevant formalism in order to collect necessary
formulas. As in B-decays, CP violating contributions can affect charm 
transitions in three distinct ways~\cite{BaBar}.

% ITEM 1
(1) CP violation can affect $D$ decay amplitudes. This type 
of CP violation (also called ``direct CP-violation'') occurs when the absolute 
value of the decay amplitude $A_f$ for $D$ to decay to a final state $f$
is different from the one of corresponding CP-conjugated amplitude, i.e.
$\left|A_f\right|\neq \left|\overline A_{\bar f}\right|$.
This type of CP-violation occurs in both charged and neutral $D$-decays and 
is induced by $\Delta C=1$ effective operators. The easiest way of observing 
it is by examining the rate asymmetries,
\begin{equation} \label{DCP}
A_{CP}(f) = \frac{\Gamma_f-\overline\Gamma_{\bar f}}
{\Gamma_f+\overline\Gamma_{\bar f}}=
\frac{1-\left|\overline A_{\bar f}/A_f\right|^2}
{1+\left|\overline A_{\bar f}/A_f\right|^2},
\end{equation}
where $\Gamma_i$ represents the $D^0 \to i$ decay rate, while 
$\overline \Gamma_i$ represents the $\barD \to i$ rate. 
A two-component decay amplitude with weak and strong phase differences is
required for this type of CP violation,
\begin{equation} \label{weakDA}
A_f = {\cal A}_f \left[
1 + r_f e^{i\left(\Delta_f+\theta\right)} \right].
\end{equation}
Here $\Delta_f$ corresponds to the strong phase difference and 
$\theta$ corresponds to the weak phase difference 
between the CP-conserving (${\cal A}_f$) and CP-violating parts of the 
decay amplitude and $r_f$ represents the (small) ratio of their absolute 
values. Note that Eq.~(\ref{weakDA}) is a scale and scheme-independent 
way to write a non-leptonic decay amplitude. While no reliable model-independent 
predictions exist for the $\Delta_f$, it is believed that
it could be quite large due to the abundance of light-quark resonances in 
the vicinity of the $D$-meson mass inducing large final-state interaction (FSI) 
phases. As the most optimistic model-dependent estimates put the Standard Model 
predictions for the asymmetry $A_{CP} < 0.1\%$~\cite{Buccella:1996uy}, 
an observation of any CP-violating signal in the current round of experiments 
will be a sign of new physics. Current FOCUS, CLEO, and Belle/BaBar results 
put rather stringent bounds on $A_{CP}$. For example, for a state $K^+ K^-$ the 
direct CP asymmetry is~\cite{Csorna:2001ww}
\bea
A_{CP} (K^+ K^-) = (0.0 \pm 2.2 \pm 0.8)\%.
\eea
Searches for this type of CP-violation in neutral charm decays 
must deal with the tagged data sample, which means that only 
$D$-mesons tagged at production, typically in $D^* \to D^0 \pi$ decay,
could be used in this analysis. This imposes restrictions on the available 
dataset of $D$'s which could be used for this analysis. 

% ITEM 2
(2) CP violation can affect the $\DDbar$ mixing matrix. This type of CP violation
occurs when effective operators that change $D^0$ into $\barD$ states, i.e. 
generate the mass and width splittings for the mass eigenstates of the 
$D^0-\barD$ system, have CP-violating coefficients. This results 
in the mixing of flavor eigenstates into the mass eigenstates,
\begin{equation} \label{definition1}
| D_{^1_2} \rangle = 
p | D^0 \rangle \pm q | \bar D^0 \rangle.
\end{equation}
Here the complex parameters $p$ and $q$ are the off-diagonal 
elements in the phenomenological parametrization of the 
$D^0-\barD$ mass matrix~\cite{Donoghue:1992dd}
\begin{equation} \label{mixmatr}
\left [ M - i \frac{\Gamma}{2} \right ]_{ij} =\left( \begin{array}{cc} ~A~ & ~p^2~\cr
~q^2~ & ~A~ \end{array} \right).
\end{equation}
CP violation in the mixing matrix occurs when
\begin{equation}
R_m^2=\left|\frac{q}{p}\right|^2=
\frac{2 M_{12}^*-i \Gamma_{12}^*}{2 M_{12}-i \Gamma_{12}} \neq 1.
\end{equation}
It is convenient to define two dimensionless variables 
$x$ and $y$ which are the normalized mass and width differences of $D_1$ and $D_2$,
\begin{eqnarray} \label{definition}
x \equiv \frac{m_2-m_1}{\Gamma}, ~~
y \equiv \frac{\Gamma_2 - \Gamma_1}{2 \Gamma},
\end{eqnarray}
where $m_{1,2}$ and $\Gamma_{1,2}$ are the masses and 
widths of the corresponding mass eigenstates, i.e. the
eigenvalues of the mixing matrix in Eq.~(\ref{mixmatr}) and
$\Gamma=\left(\Gamma_1+\Gamma_2\right)/2$.

This type of CP violation in charm can be observed most cleanly by looking for 
semileptonic decay asymmetries $A_{SL}$, like the one in Eq.(\ref{DCP}) 
with $f=X \ell\nu$. It is easy to see that $A_{SL}=(1-R_m^2)/(1+R_m^2)$~\cite{BaBar}.
This asymmetry is expected to be tiny in both the SM and many of its extensions.

%ITEM 3
(3) CP violation can occur in the interference of decays with and 
without mixing. This type of CP violation is possible for a subset of final 
states to which both $D^0$ and $\barD$ can decay. It is usually associated 
with the relative phase between mixing and decay contributions. It 
can be studied by examining the time-dependent version of rate 
asymmetry $A_{CP}(f)$ of Eq.~(\ref{DCP}).

Studies of $D^0-\barD$ oscillations offer a convenient probe of CP violation in 
the charm system. Using the notations of \cite{Bergmann:2000id}, let us write 
the time-dependent decay rates of $D^0$ and $\barD$ to a given final state $f$.
Since $x,y\ll 1$ \cite{ShD,Golowich:1998pz} and denoting ${\cal T} = \Gamma t$,
\begin{widetext}
\bea \label{timeDep}
\Gamma_f (t) &=& R_m^2 \left | \overline A_f\right|^2 e^{-{\cal T}}
\left[
~\left|\lambda_f^{-1}\right|^2 +
\left[
Re \left( \lambda_f^{-1}\right) y + Im \left(\lambda_f^{-1}\right) x 
\right] {\cal T} +
\left[
\left(x^2 + y^2 \right) -
\left(x^2 - y^2 \right)
\left|\lambda_f^{-1}\right|^2 \right] \frac{{\cal T}^2}{4}
\right],
\nonumber \\
\overline \Gamma_f (t) &=& R_m^{-2} \left | A_f\right|^2 e^{-{\cal T}}
\left[
~\left|\lambda_f\right|^2 +
\left[
Re \left( \lambda_f\right) y + Im \left(\lambda_f\right) x 
\right] {\cal T} +
\left[
\left(x^2 + y^2 \right) -
\left(x^2 - y^2 \right)
\left|\lambda_f\right|^2 \right] \frac{{\cal T}^2}{4}
\right],
\eea
\end{widetext}
where for a given final state $f$, CP violation is parametrized by  
\bea\label{Lambda}
\lambda_f^{-1} ~&=&~ \frac{p}{q} \frac{A_f}{{\overline A}_f}=
\sqrt{R} R_m^{-1} e^{-i(\delta+\phi)}+{\cal O}(r_f),
\nonumber \\
\lambda_{\overline f} ~~&=&~ \frac{q}{p} \frac{{\overline A}_{\overline f}}
{A_{\overline f}}=\sqrt{R} R_m e^{-i(\delta-\phi)}+{\cal O}(r_{\overline f}),
\eea
where $A_f$ and ${\overline A}_f$ are the amplitudes for $D^0 \to f$ and 
$\barD \to f$ transitions respectively, $\delta$ is the strong phase 
difference between $A_f$ and ${\overline A}_f$, and 
$R = \left|{{A_f}/{\overline A}_f}\right|^2$.
Here $\phi$ represents the convention-independent CP-violating weak phase 
difference between the ratio of decay amplitudes and the mixing matrix.
The corresponding expressions for $\Gamma_{\overline f} (t)$ and 
$\overline \Gamma_{\overline f} (t)$ can be found by substituting 
$f \to \overline f$ in Eqs.~(\ref{timeDep}). 

Eqs.~(\ref{timeDep}) give the most general expression for 
time-dependent decay rate up to ${\cal O}(x^2,y^2)$. However,
parameters of these equations (i.e. coefficients of $x$ and $y$
to the appropriate power) scale differently for 
singly Cabbibo suppressed (SCS) or doubly Cabbibo suppressed
(DCS) decays. For instance, $R \sim {\cal O}(\lambda^4)$  
for DCS decays like $D^0 \to K^+ \pi^-$, while $R \sim {\cal O}(1)$ for
SCS transitions such as $D^0 \to \pi^+ \rho^-$. This implies that in the studies 
of a particular DCS or SCS transition, some of the terms in Eqs.~(\ref{timeDep})
could be neglected. CP-violating parameters could be extracted by
comparing the time-dependent rates of $D^0$ and $\barD$ decays
in Eq.~(\ref{timeDep}).

%%%%%%%%%%%%%%%%%%%%%%%%%%%%%%%%%%%%%%%%%%%%%%
\section{Untagged signals of CP violation}

The existing experimental constraints~\cite{Csorna:2001ww} demonstrate that 
CP-violating parameters are quite small in the charm sector, regardless of 
whether they are produced by the Standard Model mechanisms or by some new physics 
contributions. In that respect, it is important to maximally exploit
the available statistics. It is easy to see that the rate asymmetries of 
Eq.~(\ref{DCP}) require tagging of the initial state with the consequent 
reduction of the dataset.

Another way of looking for CP-violation involves methods which are being discussed 
in connection with CLEO-c tau-charm factory measurements. They do not require
initial state flavor tagging but rely on the fact that at the threshold charm factory
initial $D^0-\barD$ state is prepared in the state with definite CP. 
An observation of a final state of the opposite CP would 
automatically imply CP-violation. These signals were discussed 
in~\cite{Bigi:1986dp}. Since they involve CP-violating decay {\it rates}, 
these observables are of {\it second order} in the small CP-violating 
parameters, a challenging measurement.

We propose a method that both does not require flavor or CP-tagging of the 
initial state and results in the observable that is {\it first order} 
in CP violating parameters. Let's concentrate on the decays of $D$-mesons 
to final states that are common for $D^0$ and $\barD$. If the initial 
state is not tagged the quantities that one can easily measure 
are the sums 
\begin{equation} 
\Sigma_i=\Gamma_i(t)+{\overline \Gamma}_i(t) 
\end{equation}
for $i=f$ and ${\overline f}$.
A CP-odd observable which can be formed out of $\Sigma_i$ 
is the asymmetry
\begin{equation} \label{TotAsym}
A_{CP}^U (f,t) =  
\frac{\Sigma_f - \Sigma_{\overline f}}{\Sigma_f + \Sigma_{\overline f}}
\equiv \frac{N(t)}{D(t)}.
\end{equation}
We shall consider both time-dependent and time-integrated versions 
of the asymmetry (\ref{TotAsym}). Note that this asymmetry does not 
require quantum coherence of the initial state and therefore is accessible in 
any D-physics experiment. From Eq.~(\ref{timeDep}) 
it is expected that the numerator and denominator of Eq.~(\ref{TotAsym}) would 
have the form,
\bea \label{NumDenum}
N(t) &=& \Sigma_f - \Sigma_{\overline f} 
= ~e^{-{\cal T}} \left[A + B {\cal T} + C {\cal T}^2 \right],
\nonumber \\
D(t) &=& ~e^{-{\cal T}} \left[ 
\left | A_f \right|^2 +  \left | \overline A_{\overline f} \right|^2 +
\left | A_{\overline f} \right|^2 + \left | \overline A_f \right|^2
%\Gamma_f + \overline \Gamma_f + \Gamma_{\overline f} + \overline \Gamma_{\overline f}
\right].~~~~
\eea
Integrating the numerator and denominator of Eq.~(\ref{TotAsym}) over time 
yields
\beq
A_{CP}^U (f) = \frac{1}{D}\left[A + B + 2 C\right],
% {\left | A_f \right|^2 +  \left | \overline A_{\overline f} \right|^2 +
% \left | A_{\overline f} \right|^2 + \left | \overline A_f \right|^2},
%{\Gamma_f + \overline \Gamma_f + \Gamma_{\overline f} + \overline \Gamma_{\overline f}},
\eeq
where $D=\Gamma \int_0^\infty dt ~D(t)$. 

Both time-dependent and time-integrated asymmetries depend on the same parameters
$A, B$, and $C$. Since CP-violation in the mixing matrix 
is expected to be small, we follow~\cite{Bergmann:2000id} and expand
$R_m^{\pm 2} = 1 \pm A_m$. The result is
\begin{widetext}
\bea\label{Coefficients}
A ~&=& ~\left (\left | A_f \right|^2 -  \left | \overline A_{\overline f} \right|^2 \right)-
\left (\left | A_{\overline f} \right|^2 -  \left | \overline A_f \right|^2 \right) =
\left | A_f \right|^2 \left[
\left (1-{\left | \overline A_{\overline f} \right|^2}/{\left | A_f \right|^2} \right)
+ R \left (1-{\left | A_{\overline f} \right|^2}/{\left | \overline A_f \right|^2} 
\right) \right],
\nonumber \\
B ~&=& ~
-2 y \sqrt{R} ~\left[ \sin\phi \sin \delta 
\left(
\left | \overline A_f \right|^2 + \left | A_{\overline f} \right|^2
\right) - 
\cos\phi \cos \delta
\left(
\left | \overline A_f \right|^2 - \left | A_{\overline f} \right|^2
\right)
\right] + {\cal O}(A_m x, r_f x, ...),
\\
C ~&=& ~\frac{x^2}{2}
\left [
\left (\left | A_f \right|^2 -  \left | \overline A_{\overline f} \right|^2 \right)-
\left (\left | A_{\overline f} \right|^2 -  \left | \overline A_f \right|^2 \right)
\right] = \frac{x^2}{2} A + {\cal O}(A_m x^2, A_m y^2).
\nonumber
\eea
\end{widetext}
We neglect small corrections of the order of ${\cal O}(A_m x, r_f x, ...)$ and
higher. It follows that Eq.~(\ref{Coefficients}) receives contributions from both 
direct and indirect CP-violating amplitudes. Those contributions have different time
dependence and can be separated either by time-dependent analysis of Eq.~(\ref{TotAsym}) 
or by the ``designer'' choice of the final state. Note that this asymmetry is manifestly 
{\it first} order in CP-violating parameters.

In Eq.~(\ref{Coefficients}), non-zero value of the coefficient $A$ is an indication 
of direct CP violation. This term might be important for SCS decays. 
The coefficient $B$ gives a combination of a contribution of CP violation in the 
interference of the decays with and without mixing (first term) and direct CP 
violation (second term). Those contributions can be separated by considering 
DCS decays, such as $D \to K^{(*)} \pi$ or $D \to K^{(*)} \rho$, where direct CP violation 
is not expected to enter. The coefficient $C$ represents a contribution of 
CP-violation in the decay amplitudes after mixing. It is negligibly small in the SM and 
all models of new physics constrained by the experimental data. Note that the effect of 
CP-violation in the mixing matrix on $A$, $B$, and $C$ is always subleading.

Eq.~(\ref{Coefficients}) is completely general and is true for both DCS and SCS 
transitions. For an experimentally interesting DCS decay $D^0 \to K^+ \pi^-$ we can neglect 
direct CP violation and obtain a much simpler expression,
\bea \label{KpiTime}
A ~&=& ~0, \qquad C ~=~ 0\nonumber \\
B ~&=& ~- 2 y \sin\delta \sin\phi ~\sqrt{R} ~
\left | \overline A_{K^+ \pi^-} \right|^2.
\eea
This asymmetry is zero in the flavor $SU(3)_F$ symmetry limit, where
$\delta = 0$~\cite{Wolfenstein:1985ft}. Since $SU(3)_F$ is badly broken 
in $D$-decays, large values of $\sin\delta$~\cite{Falk:1999ts} are possible. 
At any rate, regardless of the theoretical estimates, this strong phase could be 
measured at CLEO-c~\cite{Silva:2000bd}.
It is also easy to obtain the time-integrated asymmetry for $K\pi$. Neglecting small
subleading terms of ${\cal O}(\lambda^4)$ in both numerator and 
denominator we obtain
\beq \label{KpiIntegrated}
A_{CP}^U (K\pi) = - y \sin \delta \sin \phi \sqrt{R}. 
\eeq
It is important to note that both time-dependent and time-integrated asymmetries
of Eqs.~(\ref{KpiTime}) and (\ref{KpiIntegrated}) are independent of predictions
of hadronic parameters, as both $\delta$ and $R$ are experimentally determined 
quantities and could be used for model-independent extraction of CP-violating phase 
$\phi$. Assuming $R \sim 0.4\%$ and $\delta \sim 40^o$~\cite{Falk:1999ts} and
$y \sim 1\%$~\cite{Golowich:1998pz} one obtains 
$\left| A_{CP}^U (K\pi) \right| \sim \left(0.04\%\right)\sin\phi$.
Thus, one possible challenge of the analysis of the asymmetry 
Eq.~(\ref{KpiIntegrated}), is that it involves a difference of two large 
rates, $\Sigma_{K^+\pi^-}$ and $\Sigma_{K^-\pi^+}$, which should be
measured with the sufficient precision to be sensitive to $A_{CP}^U$, 
a problem tackled in determinations of tagged asymmetries in $D \to K \pi$ 
transitions.

Alternatively, one can study SCS modes, where $R \sim 1$, so the 
resulting asymmetry could be ${\cal O}(1\%)\sin\phi$. However, the final states 
must be chosen such that $A_{CP}^U$ is not trivially zero. For example, 
decays of $D$ into the final states that are CP-eigenstates would result 
in zero asymmetry (as $\Gamma_f=\Gamma_{\overline f}$ for those final states) 
while decays to final states like $K^+ K^{*-}$ or $\rho^+ \pi^-$ would not. 

The final state $f$ can also be a multiparticle state. In that case,
more untagged CP-violating observables could be constructed.  
For instance, three body decays can exhibit CP-violating Dalitz plot 
asymmetries, like the $E_+ \leftrightarrow E_-$ asymmetry of the 
Dalitz plot in the decay $D \to K_S \pi^+(E_+) \pi^-(E_-)$. 
Similar studies of Dalitz plot asymmetries in $B_d$-decays were suggested 
in~\cite{Burdman:1992vt}~\footnote{That study, however, neglected
the contribution due to lifetime difference $y$ (which a good approximation
in $B_d$ decays), leading for $D$-meson transitions.}. 
Untagged studies of Dalitz plot population asymmetries resulting from 
the enantiometric intermediate states were proposed in~\cite{Gardner:2003su} 
to study direct CP-violation in $B_d$ decays.
The use of untagged samples were also proposed to measure $\Delta\Gamma$ and 
CKM phase $\gamma$ in $B_s$ decays~\cite{Dunietz:1995cp}~\footnote{I thank 
Y.~Grossman for pointing this reference to me.}, where large
values of $x_s$ cause these terms to oscillate rapidly 
and, effectively to cancel out for sufficiently large $t$. The 
situation in charm transitions is exactly opposite.

As any rate asymmetry, Eq.~(\ref{TotAsym}) requires either a ``symmetric'' 
production of $D^0$ and $\barD$, a condition which is automatically satisfied 
by all $p\overline p$ and $e^+ e^-$ colliders, or a correction for 
$D^0/\barD$ production asymmetry.

%%%%%%%%%%%%%%%%%%%%%%%%%%%%%%%%%%%%%%%%%%%%%%
\section{Conclusions}

We propose a method of searching for CP-violation in charm decays which does 
not require either flavor or CP tagging of the initial state. The resulting 
asymmetry is first order in CP-violating parameters, which is important for 
charm transitions. The unique feature of the asymmetry of Eq.~(\ref{TotAsym})
is that, apart from the small CP-violating phase, it could be sizable even for 
two body final states. This occurs because of the large $SU(3)_F$ symmetry 
breaking in charm transitions.

This work was supported in part by the U.S.\ National Science Foundation
under Grant PHY--0244853, and by the U.S.\ Department of Energy under
Contract DE-FG02-96ER41005. I thank Dave Cinabro, Susan Gardner, Yuval Grossman, 
Rob Harr, and Jure Zupan for useful discussions.

%%%%%%%%%%%%%%%%%%%%%%%%%%%%%%%%%%%%%%%%%%%%%%
%\begin{references}

\end{document}